\documentstyle[prd,aps]{revtex}
\begin{document}
\draft
\preprint{
UMD-PP-98-
{}~hep-ph yymmnn}
\title{ Possible Manifestation of Heavy Stable Colored Particles in
Cosmology and Cosmic Rays}
\author{R. N. Mohapatra$^{1}$ and
S. Nussinov$^{2}$}
\address{$^{(1)}${ Department of Physics, University of Maryland, 
College Park, MD-20742, USA.}}
\address{$^{(2)}${ School of Physics and Astronomy, Tel Aviv University,
Tel Aviv, Israel\\
 and Department of Physics, University of South Carolina, Columbia, SC-29208}}
\date{August, 1997}
\maketitle
\begin{abstract}

We discuss the cosmological implications as well as possible observability
of massive stable colored particles which often appear in the discussion of
physics beyond the standard model. We show that if their masses are bigger
than few hundred GeV and they saturate the halo densityand/or they 
occur with closure density of the universe, they can
be ruled out by existing limits on heavy stable particles from analysis
of anomalously heavy isotopes of ordinary nuclei as well as from the
observations in WIMP detectors. We also derive constraints on their masses
if they annihilate to produce gamma rays. We then comment on the
possibility that these particles could be responsible for the ultra high
energy (UHE) cosmic rays with energies $\geq 10^{20}$ eV observed in several
recent experiments and in particular point out that their low inelasticity
argues against the possibility they can explain the observed UHE events.
           
\end{abstract}

\pacs{ \hskip 4 cm  UMD-PP-98-17}

\section{Introduction}
 Heavy stable particles carrying gauge charges naturally arise in many
field theoretic contexts. Such particles with electric charges have been
extensively discussed in the past\cite{1} in connection with their possible
role as dark matter. We will assume in what follows that
these particles are electrically neutral but carry color. 
Examples of such particles (to be denoted henceforth by $X$) 
can arise in supersymmetric models where
the gluino ( rather than the photino ) is the lightest
SUSY particle in which case we will have stable $\tilde{g}g$ bound states
with masses in the low TeV range (or much lower masses in the light gluino
scenario\cite{farrar1}). There can be examples of gauge mediated
supersymmetry breaking models\cite{satya} where one can have color octet
messengers that can be stable and heavy (with masses easily in the 10 TeV
range). There are also examples of mirror models  for particle
where the particle spectrum as well as the gauge symmetries are completely
doubled\cite{bere}. These models have been proposed in connection with
understanding the neutrino puzzles or to solve the strong CP problem.
If in these models the color $SU(3)_c$ groups of the mirror universe
mixes with the that of the known universe, then the mirror quarks will
share the strong interaction but not the observed electroweak interactions.
Finally, the well known magnetic monopoles 
that arise in grand unified theories, although overall color neutral
can carry a color cloud in addition
to magnetic charge if the initial GUT group is simple. 
 
In this paper, we comment on the cosmological aspects and possible experimental
observability of such particles\footnote{For an earlier
analysis of strongly interacting particles which constitute the halo dark
matter see G. Starkman et al.\cite{dimo}. Our work is
complementary to this  and addresses several issues not covered in
Ref.\cite{dimo}}. We use a plausible ansatz for the annihilation 
as well as scattering cross-section for these particles and draw
the following conclusions: (i) if their masses are above 
several hundred GeV, and  they saturate the halo density,
they should have been seen in the existing underground detectors
 and are excluded by present searches\cite{caldwell,avig};
and secondly (ii) in the same mass range they are also excluded by
 the recent limits on anomalously heavy isotopes of ordinary 
nuclei\cite{hemmick}. Furthermore, if either the $X$ particles are 
their own antiparticles (as in the case of color octets or gluino-gluon
bound states) or the $X$ and $\bar{X}$ are equally abundant
( saturating the halo density in both the cases), 
their annihilation in the halo can lead to energetic gamma ray
fluxes. Present limits on these fluxes rule out the mass range 
$M_X\leq 300$ TeV.

We also focus on recent suggestions in the 
literature to utilize particles\cite{kuzmin,farrar} in the above
general category to explain the puzzling phenomena of ultra high energy (UHE)
cosmic ray events\cite{nagano} observed in several independent experiments
in the past decade. One could imagine these colored but
electrically neutral particles\cite{farrar} playing the role of the
original high energy particle as the cause of these UHE cosmic rays.
 Our discussion leading to the above constraints also makes it clear
that if these particles have
masses above several hundred GeV, their hadronic collisions 
have very low inelasticity making them unsuitable as the UHE primaries
which need to have significant inelasticities in order
 to generate the observed shower characteristics.
                                                          
This paper is organized as follows: in section 2, we present estimates of
the annihilation, scattering and pair creation cross-sections of such
particle. Using these estimates in sec. 3, we proceed to discuss
the residual abundance of such particles in the present universe, which
we find, could provide the closure density of the universe. In section
4, we discuss their
prospects for nuclear capture and formation of rare isotopes
and their observability in various experiments.
 We then comment briefly on the connection between the UHE primaries
and the massive $X$- particles in section 5. In section 6, we   
discuss in some detail the suggestion\cite{weil} that the
magnetic monopoles are responsible for
the UHE cosmic ray events and argue that very low inelasticity
of monopole-nuclear collisions tend to preclude this mechanism.

\section{Reaction cross-sections for the $X$ particles}

Because the $X$ particles are much more massive than the usual hadrons,
we must assume that their mass originates not from QCD but
from physics at some new scale such as 
the scale responsible for breaking supersymmetry.
Consequently we envision a small core for the $X$ particle, of size 
$~O(M^{-1}_X)$  where the bulk of the mass is residing. 
This specific novel feature distinguishes
the $X$ particles from the ordinary hadrons.

In $X$-nuclear or $XX$ low momentum transfer collisions, only gluonic
degrees of freedom residing in a hadronic cloud of size $\Lambda^{-1}_{QCD}$
$\simeq Fermi$ need to be involved. Hence we expect to have generic
hadronic cross-sections
\begin{eqnarray}
\sigma_{X-N}\approx \Lambda^{-2}_{QCD} \simeq 10~~mb
\end{eqnarray}
The same estimate applies also to $X-X$ elastic cross-section. On the other
hand in $X-\bar{X}$ annihilation process, the cores of $X$ and $\bar{X}$ 
particles should overlap and we therefore expect this cross-section to
be much smaller:
\begin{eqnarray}
v\sigma^{ann}_{X-\bar{X}} \approx \frac{\alpha^2_s}{M^2_X}\approx
\alpha^2_{s}(10^{-8}-10^{-10})~GeV^{-2}
\end{eqnarray}
Likewise, just as for $t\bar{t}$ pair production, $X-\bar{X}$ production
cross-section in hadronic collision is also very small:
\begin{eqnarray}
\sigma_{hadrons\rightarrow X\bar{X}}\approx
\frac{\alpha^2_s}{M^2_X}f(\frac{M_X}{\sqrt{s}})
\end{eqnarray}
with $f(\frac{M_X}{\sqrt{s}})$ representing some further (structure function)
suppression. 

The above crude estimates of the cross-sections for processes
involving the $X$ particles suffice for our discussion and unless the above
intuitive arguments are completely off, the constraints obtained for the
$X$ particles will remain qualitatively unchanged.

One may wonder whether the interactions of the soft hadronic clouds could
modify the $X-\bar{X}$ annihilation cross-section.  
The soft hadronic interction extends over typical hadronic
scale ( say, a Fermi) and also have typical hadronic strength i.e. effective
potential depth of order $U_0\leq GeV$. 
If this interaction can pull together
the ``cores'' of the two colliding $X$ particles,
then one would expect the $\sigma^{ann}_{X-\bar{X}}$ to be of the same order
as the $\sigma^{el}_{X-X}$ i.e. of order $\sigma_{hadron}\approx (Fermi)^2$
rather than the small value (Eq.2). The 
hadronic cloud interaction is strong enough
to pull the ``cores'' of the $X$ particles
when the $X$ particles are sufficiently slow moving,
$\beta_X\leq \sqrt{\frac{GeV}{M_X}}$,
with their center of mass kinetic energy
less than a GeV. In this case we would
expect a significant enhancement in their annihilation cross-section
all the way to the value of say nucleon-antinucleon annihilation at
low velocities:
\begin{eqnarray}
\sigma^{ann}_{X-\bar{X}}\approx \frac{1}{\beta} (Fermi)^2
\end{eqnarray}
This estimate will hold for example for relic $X$ particles in the late
cool stages of the universe. However the earlier estimate Eq. 2 will hold in the
early hot stage when the center of mass kinetic energy 
of the collision exceeds $U_0\approx \Lambda_{QCD}$ and we can ignore the
cloud effect. We will see in the next section that this
result has a profound implication for the number density of the relic
$X$ particles and their subsequent behaviour.

As we will see in the next section, the freezeout temperature $T^*$
depends only logarithmically on $\sigma^{ann}_{X-\bar{X}}$. The ratio
$T^*/M_X$ is always larger than $0.02$ so long as $\sigma^{ann}\leq (Fermi)^2$.
For the value of $M_X\sim 1-10$ TeV (as is the case we are considering),
the temperature at decoupling $T^*\approx 20-200$ GeV, 
is considerably larger than
$\Lambda_{QCD}$ making the enhancement of the cross-section noted above
irrelevant for the relic abundance.

\bigskip

\section{Relic density of $X$ particles}

In the very universe $T\gg M_X$, we expect the density of $X$ particles to be
comparable to that of other relativistic particles. However as the universe
cools below $M_X$, the annihilation of the $X-\bar{X}$ pairs  
dominates over their production gradually decreasing the ratio 
$n_X/n_{\gamma}$, until the freezeout temperature where rate of
annihilation becomes slower than the expansion rate of the universe.
 The annihilation rate
is given by $\frac{dn_X/dt}{n_X}\equiv\sigma^{ann} v n_{X}$ so that
 the freezeout temperature is dictated by the following inequality\cite{kolbt}:
\begin{eqnarray}
\frac{\alpha^2_s}{\pi M^2_X}\leq \frac{1.66\sqrt{g^*}T^2}{M_{P\ell}}
\end{eqnarray}
Using next $n_X\approx g_X (\frac{M_X T}{2\pi})^{3/2} e^{-\frac{M_X}{T}}$, 
(where $g_X$ is the number of degrees of freedom of the $X$ particle: e.g.
for a color octet spin zero field, $g_X=8$ etc)
we can rewrite
(Eq. ) in terms of the dimensionless quantity $\xi\equiv \frac{M_X}{T}$ 
as follows:
\begin{eqnarray}
\frac{2g_X}{(2\pi)^{5/2}}\alpha^2_s e^{-\xi} = 
\sqrt{g^*} \xi^{-1/2}\frac{M_X}{M_{P\ell}}
\end{eqnarray}
Taking $M_X/M_{P\ell}= 10^{-16}-10^{-15}$, $g^*\simeq 10$ and $\alpha_s\simeq
.12$, we find 
\begin{eqnarray}
\xi\approx 1/2 \xi + ln 10^{14} + ln\frac{2 g_X\alpha^2_s}{(2\pi)^{5/2}}
\approx 30
\end{eqnarray}
Finally using $n_{\gamma}\approx g^* T^3 $ we have
\begin{eqnarray}
\frac{n_X}{n_{\gamma}}= \frac{1.66\pi^3}{2.4\sqrt{g^*}\alpha^2_s}
\xi \frac{M_X}{M_{P\ell}}
\approx 170 \xi \frac{M_X}{M_{P\ell}}\approx 5\times (10^{-12}-10^{-13})
\end{eqnarray}
Using next $\eta_B\equiv \frac{n_B}{n_{\gamma}}\approx 6\times 10^{-10}$, we
find
\begin{eqnarray}
\frac{n_X}{n_{B}}\simeq  10^{-2}- 10^{-3}
\end{eqnarray}
Hence the mass densities are in the ratio
\begin{eqnarray}
\frac{\rho_X}{\rho_B}=\frac{n_XM_X}{n_Bm_B}=1-100~~~~~~~(for M_X\simeq 
1-10 ~TeV)
\end{eqnarray}
Thus we find the nice feature that taking $M_X$ in the range of 1-10 TeV
suggested by independent considerations naturally results in closure density
of the universe. This is of course not an entirely new result but clearly
 related to the discussion of the lightest supersymmetric particle (LSP)
 with a corresponding weak annihilation
cross-section being the dark matter of the universe.

We will now show that there are severe difficulties with such relic particles
if the $X$ have hadronic strength interactions with ordinary matter.
Once the temperature $T\leq \Lambda_{QCD}\simeq .3 $ GeV , the QCD phase
transition temperature, the gluon and quark gas disappear. At this
hadronic recombination temperature, the purely strogly interacting particles
almost decouple from the baryons (and from the rest of the background
radiation). Once $\rho_X\geq \rho_{\gamma}$, a standard cold dark matter 
scenario may be implemented: the $X$ particle would start forming structure
and galactic halos in particular. Subsequently the potential wells of these
particles will trap the dissipating baryons which form the galactic discs.
If we assume that the local halo density $\rho\approx 0.3 $ GeV/(cm)$^3$,
is generated by our $X$ particles, we will have 
\begin{eqnarray}
n_X\simeq 0.3~GeV/M_X\simeq 10^{-5}~cm^{-3}
\end{eqnarray}
This of course significantly higher than the value of relic density
mentioned earlier (using $n_{\gamma}\approx 300~cm^{-3}$ one finds
that $n^{relic}_X\simeq 10^{-10}- 10^{-11}$). We will now discuss three 
possible manifestations of the $X$- particles: (i) in underground
detectors ; (ii) via rare isotope formation and (iii) via annihilation in 
the halo. This discussion will be done for both cases of 
galactically clustered as well as cosmological, unclustered, $X$ particles.

\section{Manifestation of the $X$ particles}

\subsection{Wimp detectors and $X$ particles}

Let us first consider issue (i) - namely 
the possible manifestation of the $X$
particles in underground detectors if the $X$'s saturate the halo
density. Even though the $X$ particles
are strongly interacting (unlike the usual wimp candidates for dark matter)
their penetration depth in earth or water can vastly exceed that of the
normal hadrons. This is so despite the fact that 
as indicated in sec. II, the typical $X$-nucleus
scattering may well be of the size of normal hadronic cross-section- i.e.
$\sigma \approx \pi R^2(A,Z)$ with $R\simeq 1.2 A^{1/3}$ Fermi, the nuclear
radius. The reason for this is the large mass mismatch between the
$X$ particle and a normal hadron. This implies that we require many
collisions 
\begin{eqnarray}
n_{coll}\approx M_X/ <m_{(A,Z)}>\simeq \frac{M_X}{20~GeV}
\end{eqnarray}
(where we have chosen
$<m_{(A,Z)}>\approx 20 $ GeV for average water-air-crust nuclei) 
before the initial $X$ particle appreciably 
slows down or is deflected from its path. The effective penetration 
depth is then given by 
\begin{eqnarray}
\ell_{p} = \lambda n_{coll}
\end{eqnarray}
 where $\lambda$ is the 
mean free path between succesive collisions given by
\begin{eqnarray}
\lambda = \frac{1}{n(A,Z) \sigma(X-(A,Z))}\approx \frac{10^{3}~cm}{\sigma
/(Fermi)^2}
\end{eqnarray}
where we used $n(A,Z)= N_{Avagadro}\rho/ <m(A,Z)>$ and $\rho\approx 3~gm~
cm^{-3}$, the average density of the crust of the earth.
As a result the penetration length is given by
$\ell_p=\lambda \frac{M_X}{20~GeV}\approx \frac{10^3~cm}{\sigma/(Fermi)^2}
\left(\frac{M_X}{20~GeV}\right)$. Thus for $M_X\geq 2$ TeV, the penetration
depth is about a kilometer which is the typical depth for most underground
Wimp detectors. Let us emphasize again that
the key to the large number of the collisions, leading to large penetration
depth is the small energy loss in any given X-nucleus collision due to the
large mass of the $X$ particle. Even if the collision "drags" the nucleus
to the same velocity $\beta_X$ of the $X$ particle, the kinetic energy
of the nucleus will only be $\frac{1}{2} m(A,Z) \beta^2_X$ i.e. a factor
$m(A,Z)/M_X$ smaller than the kinetic energy of the $X$ particle. Hence a large
number of collisions $n_{coll}=M_X/m(A,Z)$ will be required to reduce its
energy by a factor $e$ (i.e. $(1-\frac{1}{n_{c0ll}})^{n_{coll}}\equiv e^{-1}$).

A somewhat more careful estimate of the $X-$nucleus cross-section 
$\sigma(X-(A,Z))$ can be obtained as follows.
It has been argued before\cite{nussinov} that the slow $XG$
(where $G$ stands for the gluon) color neutral bound state, to which
we will continue to refer to as the $X$ particle, sees 
the nucleus as a potential
well of radius $R(A)$ and depth of order $V_0\simeq 10 $ MeV. The
scattering is predominantly in S-wave. For the relative motion with reduced mass
$m\simeq m(A,Z)$, energy $\approx \frac{m(A,Z)\beta^2}{2}\approx 10 $ keV
and momentum $p\simeq m(A,Z)\beta\approx 20 $ MeV (assuming virial velocities
for the particles) one gets a phase shift
\begin{eqnarray}
\delta_0 \simeq -pR + arctan\left(\frac{p}{\sqrt{p^2+2mV_0}}
tan\sqrt{p^2+2mV_0}R\right)\approx -pR
\end{eqnarray}
where we have used the fact that $\sqrt{2mV_0} R\gg 1$ so that the cross-
section is essentially geometric i.e.
\begin{eqnarray}
\sigma \approx \pi R^2 \approx \pi A^{2/3} (Fermi)^2
\end{eqnarray}
Using this one gets via Eq. 13, 14 and 16 a 
penetration depth $\ell_p\approx 100$ meters 
in the rock for $M_X\approx 2$ TeV, confirming the earlier estimate.

If the above analysis applies, then 
the "wimp" detectors would be able to detect
the full flux of the $X$ particles if located at appropriate depths where
the enrgy attenuation is very small. Whereas most underground detectors
(e.g.\cite{avig}) are located at greater depths, some are relatively shallow
(\cite{caldwell}) and will suited for this search. Note further that since the
$X-$nuclear cross-sections are about $\approx 10^{12}$ times bigger than
the generic heavy Dirac neutrino cross-section, if the local $X$ particle 
density is either given by the estimated halo density above or even the
much smaller relic density, the detectors that ruled out the heavy Dirac
 neutrinos would a-fortiori also rule out the heavy $X$ particle.

We should note that the $X$ particles can be captured in nuclei 
once the center of mass collision kinetic energy 
$\frac{1}{2}M_X\beta^2$  is smaller than the expected few MeV X-nuclear
binding energy i.e. $\beta_X\leq \sqrt{\frac{10^{-2}~GeV}{m(A,Z)}}
\approx \frac{1}{50}$ which is satisfied for virial velocities. However
once the nucleus is captured onto the $X$ particle, its crosssection
(hence its attenuation length) will not necessarily increase (decrease)
and in principle the nucleus could be dislodged in subsequent collisions.
Hence any possible capture of the $X$ particle need not change the
above estimate. However accompanying the capture of nucleus by the $X$,
there could be some rather striking features as $\gamma$ or neutron
emission which could serve as more sensitive signatures although it
must be remembered that extra photon emission rate will down by a factor
of $\alpha_{em}$. 
\bigskip

\subsection{Rare $X$-isotope formation}

We have estimated above that the relic cosmological abundance of $X$-paritcles
relative to hydrogen
is about $10^{-3}$ or so. Clearly this ratio need not be maintained locally
on earth where we have a much enhanced baryon number density. To estimate
the terrestrial $X$ particle effect, we first note that a flux $\Phi_X$ of
such particles impinging the earth thru Hubble time yields a total of
$\Phi_X. t_H\simeq 2\times 10^{17}\Phi_X$ $X$ -particles falling on each
square centimeter on the earth. These $X$ particle 
would most likely accumulate in the ocean. For an ocean depth
$\sim$ 10 kilometers,
and surface hydrogen density of $n_H= \frac{2}{18} N_{Avagadro} 10^{6}= 
10^{29}H/ cm^2$, $r_H\equiv \frac{n_X}{n_H}\approx 10^{-12}\Phi_X$ and a 
similar number obtains for oxygen also. After slowing down to thermal 
velocities the $X$-particle will be captured by the nuclei in the ocean
water- most likely the oxygen nuclei- forming anomalous heavy ($\approx M_X$)
 isotopes. A careful search for precisely such heavy isotopes have
been carried out in a recent experiment by Hemmick et al\cite{hemmick} and they
find that the upper bound on the ratio $r_{Oxygen}\leq 4\times 10^{-17}- 3
\times 10^{-14}$ and $r_{H}\leq 2\times 10^{-24}-~3 \times 10^{-20}$ 
(for $M_X = 0.1-10$ TeV). 
Even if the collection period is shortened to 100 million
years (a reasonable lifetime of the ocean), 
these upper bounds would seem to exclude $X$ particle fluxes 
$\Phi_X\geq 1~cm^{-2}~sec^{-1}$. 
If the $X$'s constitute the dark galactic halo the expected flux 
$\Phi_X$ can be estimated to be 
\begin{eqnarray}
\Phi_X\approx n_X v_X = \frac{0.3~GeV}{M_X/~GeV} 3\times 10^7\simeq 10^{3}-
10^{4}~~~for~~ M_X=10-1 ~TeV
\end{eqnarray}
Thus $X$ as a halo dark matter is ruled out by the experiment in Ref.
\cite{hemmick}. On the other hand, if $X$-particles do not dominate
$\rho_{cosmos}$ and do not cluster in the Halos, so that Eq. 8 applies,
then $\Phi_{X}\approx n_Xv_X\approx 6\times (10^{-3}-10^{-2})$ which
given the crudeness of our estimate is marginally consistent with the above
lower bound.
\bigskip

\subsection{Annihilation in the Halo}

In this section, we consider the possibility that if the $X$ particle is
its own antiparticle or the $X$ and $\bar{X}$ are present in the halo
in equal abundance, their annihilation can give rise to energetic
$\gamma$-ray fluxes. Present bounds on such fluxes can then lead to
constraints on the $X$ particle density $n_X$. We derive this below.

We pointed out in section 2 that at small velocities $\beta \approx 10^{-3}$,
where the $X$ energies are below $\Lambda_{QCD}$, the $X-\bar{X}$ annihilation
is enhanced and is given by $\sigma_{X\bar{X}}\approx \frac{1}{\beta_X}
(Fermi)^2$. The $X$ particles in the halo will then annihilate at a rate
\begin{eqnarray}
\frac{dn_X/dt}{n_X}= n_X v_X \sigma_{X\bar{X}}\approx n_X (Fermi)^2 c
\end{eqnarray}
with $n_X\approx \frac{0.3~GeV}{M_X/GeV}\approx 10^{-4}~cm^{-3}$ for $M_X=
3$ TeV. This
implies a halo lifetime of $ 10^{20}$ sec., considerably exceeding $t_H$.
As a result, only a fraction $\approx 10^{-3}$ of the initial $X$'s will
annihilate in Hubble time. Thus any
 galaxy will lose $\approx 10^{-3}$ of its rest mass via such annihilation.
This exceeds the total energy output of stars. 

The problem however is
much more severe. The stellar radiation is largely carried away by the
optical photons. whereas roughly 40 per cent of the energy released in
$X-\bar{X}$ annihilation will convert to 100 GeV to TeV photons. There are
extremely strong bounds on these fluxes on earth\cite{kolb}:
\begin{eqnarray}
\Phi_{\gamma}\leq 10^{-5}~cm^{-2}sec^{-1}sr^{-1}
\end{eqnarray}
Since $\Phi_{\gamma} \simeq n^2_X\sigma_{X\bar{X}}v_X L$, Eq. 9 implies
$n_X\leq 10^{-6}~cm^{-3}$ where we have used $L\simeq 9$ kilopersecs
as a typical halo radius. 

In addition, these decays will also inject a large number of
relativistic $e^+e^-$ which if trapped in the galaxy will have an energy
density of $\sim 0.3f $ GeV cm$^{-3}$ where $f\simeq 10^{-3}$ is the decaying
 fraction. Using equipartition of energy, this should not exceed the
energy density in the galactic field which can be estimated to be
$u_B\simeq \frac{(3\times 10^{-6}~Gauss)^2}{8\pi}= 0.2~eV~cm^{-2}$.
However, a naive estimate of $u_{e^+e^-}$ gives $\approx 3\times 10^{-4}~GeV
\simeq 3\times 10^5$ eV which is $10^6$ times larger. Thus both these arguments 
would imply that the halo density cannot be dominated by the massive
$X$ particles. For unclustered $X$-particles, the analog of Eq. 19
with $L$ the effective halo size replaced by the Hubble radius, we find
$n_X\leq 10^{-9}$ cm$^{-3}$ (ignoring small redshift effects). Note that
this is comparable to the cosmological prediction for $n_X$ in Eq. 8. 

\bigskip

\section{Massive colored particles as the source of UHE cosmic rays}
During the past decades cosmic ray events with energies 
greater than $ 10^{20}$ eV have been observed
\cite{nagano} in several independent experiments. This may
call for new physics since
cosmic ray proton with $E\geq 5\times 10^{19}$ eV, scatter off
the cosmic microwave background producing nucleon resonances
leading to the Griesen-Zatsepin-Kuzmin cutoff.
Moreover, the
mean free path for this interaction is about 6 Mpc (or redshift 
$z=0.0125$). Since such energetic
protons are unlikely to be bent by the galactic magnetic fields
their direction will point directly to the source which must be less than
6 Mpc in distance if protons are to be the source of these UHE showers.
No such source has been found\cite{elbert}. 
On the other hand in the same general
direction, there are some viable sources at greater distances
 (e.g. the quasar 3C 147 at a distance of 240 Mpc
or the Seyfert galaxy MCG 8-11-11 at a distance of about 10 Mpc).
For the primaries to be coming from these distant sources, they must 
 avoid the analog of the GZK cutoff and 
must be strongly interacting to
produce showers\cite{farrar}. Therefore electrically neutral, colored particles
are natural candidates for the primary particles. Furthermore, in
order to successfully reproduce the observed extensive showers, these
particles should have inelasticities comparable to those of the usual
primary cosmic rays, namely protons or nuclei. In primary proton collisions,
about half of the initial energy is utilized to make the energetic
secondaries. The latter inelasticity is high. 
The arguments of the previous section implies that if the
UHE primaries were heavy ($\geq$ few hundred GeV) neutral colored
particles they would have been seen in the underground detectors. Of course,
if they are light (as in the model of Chung et al.\cite{farrar}), our
arguments do not apply. We now turn to the suggestion that
heavy monopoles of grandunified theories may explain the UHE cosmic ray
showers\cite{weil}.

\section{Monopoles as source of ultra high energy cosmic rays}

It was suggested in Ref.\cite{weil} that monopoles of grand unified
theories may be responsible for the ultra high energy (UHE)
cosmic rays with energies above $10^{20}$ eV\cite{weil}. The attractiveness
of the suggestion arises from the fact that this offers a natural mechanism
for acceleration of the particles to such ultra high energies by the galactic
magnetic fields. For natural values of the monopole charges expected in GUT
theories i.e. $E_{M}\simeq \sqrt{N} q_{M} B_G L_{coh}\simeq 10^{20}$ eV
where $q_{M}= q_e/2\alpha$ is the monopole 
magnetic charge, $N$ is the number of coherent
domains traversed and $L_{coh}$ is the typical length of a coherence domain
expected to be of order $ 100$ persecs. To mimic the observed cosmic ray
events, the monopole mass $m_M$ must be less than $10^{10}$ GeV since it must
be relativistic. There is also a kinematically a maximal value for the
inelasticity i.e. energy transferred from the monopole in the scattering 
process on a target. To obtain this the authors of Ref. 7 
note that the maximum energy transferred
in a collision of a monopole of mass $m_{M}$ and energy $E_{M}$
on a target nucleon of mass $m$ is given by:
\begin{eqnarray}
\zeta_{max} = \frac{2mE_{M}}{(2mE_{M}+m^2_{M})}
\end{eqnarray}
obtained for backward center of mass elastic scattering. We see that for
UHE case of $E_{M}\simeq 10^{12}$ GeV, $\zeta$ acquires its 
maximal value ($\zeta=1/2$) for $m_{M}\sim 10^6$ GeV and decreases for higher
 values quadratically. If this maximum value for the inelasticity were
attained, it would be very favorable for the interpretation of the
monopoles as the source of UHE's. To see if this maximal inelasticity
value can in fact obtain, one
has to consider the dynamics of the monopole collision. We find
that such high 
values for the inelasticity are unlikely to be realized in practice
if either one of the two collision mechanisms that we consider 
dominates. Below we describe these two monopole-nucleus 
collision mechanisms and estimate the reaction rates and the
 inelasticities in both cases.

\subsection{Hard electromagnetic scattering of monopoles} 

The most obvious reaction mechanism for initiating the UHE events by
the monopoles utilizes their magnetic interaction with charged particles.
The dominant mechanism briefly alluded to in Ref.\cite{weil} involve
nuclear dusruption due to the Lorentz boost generated large transverse
$E$-field seen by the stationary nucleus:
\begin{eqnarray}
E_T=\frac{\beta\gamma q_M}{2\alpha r^2}\approx \frac{\gamma q_M}{2\alpha b^2}
\end{eqnarray}
with $b$ being the impact parameter and $\gamma =\frac{E_{M}}{m_{M}}
=10^{12}/ m_{M}~in~GeV$. This field can disrupt the nucleus by
accelerating the protons (relative to the neutrons). Let us estimate
how big this effect is.

The external perturbation seen by the stationary nucleus is:
\begin{eqnarray}
H'(t) = e {\bf E}_T\cdot\Sigma {\bf r_i } f(t)
\end{eqnarray}
 where ${\bf r_i}$ is the $i$th proton's location relative to the center of 
the nucleus and $f(t)$ is a pulse shape normalized as
\begin{eqnarray}
\int f(t) dt \simeq \Delta t \simeq \frac{b}{\gamma c}
\end{eqnarray}
where $b$ is the impact parameter and the $1/\gamma$ factor is due to
squeezing towards the transverse direction of the $B$ and $E$ field lines 
of the moving monopole. This leads to a $\gamma$ independent product
$E\Delta t$. This product in turn controls both 
the boost invariant transverse momentum
imparted $\sim F\Delta t$ and the excitation probability. The probability of
exciting the nucleus is given in first order time dependent perturbation
theory by the following sum over all non-ground states:
\begin{eqnarray}
P=\Sigma P_{0\rightarrow n}\leq e q_M Z^2E^2\Sigma_{n=1}\int
|f(t) e^{i\Delta E_n t} dt|^2 |<0|{\bf r_i}|n>|^2\\
\leq (ZE)^2|\int f(t) dt|^2<0|{r}^2|0>\approx Z^2(E\Delta t)^2 \bar{R}^2\\
\approx Z^2 \bar{R}_{nucl.}^2/ b^2
\end{eqnarray}
where Schwarz inequalities, completeness and $eq_M=1$ have been used. 
Thus only if
$b\leq Z\sqrt{\bar{R}^2_{nucl.}}\approx Z\times 1.2 A^{2/3} Fermi
\approx 60 ~Fermi$,
will we have $P\simeq 1$. The excitation cross-section is therefore smaller
 than $\sigma_{dis}\approx \pi b^2\approx 10^{-22}$ cm$^2$. Furthermore,
in each disruption, the total energy loss in the excited system rest 
frame does not exceed the total nuclear binding energy i.e. $\Delta E^*\leq
100$ MeV. This gets boosted at most to $\Delta E = \gamma \Delta E^*$ in the
 laboratory frame. Even in this extreme case the fraction of
the initial monopole energy lost in GeV in a collision is
\begin{eqnarray}
\eta \leq (\gamma m_{M})^{-1}(\gamma\Delta E^*)= \Delta E^*/m_{M}=0.1~GeV/
m_{M}
\end{eqnarray}
Along the line of sight in the atmosphere 
of average length $\ell\approx 20$ km, there
are at most $N_c= \ell~ n_{Nitrogen}\sigma_{dis}\leq 10^4$
such collisions and the total energy loss is 
therefore $N_c\eta \simeq 10^3/m_{M}$, 
which is very small indeed for values of the monopole mass of interest.

Let us next consider the  "hard collision" of a point like monopole with a
 point like charge. For a classical collision at impact parameter $b$,
 the boost invariant transverse momentum transfer is given by 
$\Delta p_T= e E_T \Delta t\approx \frac{e q_{M}}{b^2} b= eq_{M}/b$,
where as discussed above $E_T = B_T \gamma\simeq eq_{M}\gamma/b^2$
and $\Delta t\approx b/\gamma$. This is precisely the same as the
momentum transfer in a hypothetical Rutherford scattering with charges 
\footnote{ It is amusing to recall
 that in nonrelativistic
classical mechanics there is an exact description of the motion of a point
charge in the field of a monopole. It is a convergent and a reflected
spiral restricted to the Poincare cone with vertex at the origin and
opening angle $\pi-\theta_0$ with $\theta_0\approx eg/L= eg/pb$, the ratio
of the electromagnetic to the orbital angular momentum, such that when the cone
is opened into a plane, the spiral becomes a zigzaging straight line\cite{XXX}.
The actual final scattering angle is $\theta\approx \theta_0$ and is just
the same as for a Coulomb problem with $Q=g_M$ at the origin. Independently of
all the above, we can compare the scattering amplitudes for charge-charge
($e-e$), monopole-monopole ($q_M-q_M$) and charge -monopole scattering
($e-q_M$) via the inequality $A_{e-e}~A_{q_M q_M}\geq A^2_{e-q_M}$.}
$Z_1=e$ and $Z_2 = q_{M}$.
   Carrying this analogy to the field theoretic
domain, we write down the Rutherford like cross-section for magnetic monopole
and charge scattering as:
\begin{eqnarray}
\frac{d\sigma}{dt}|_{q_{M},e}(t) \simeq \frac{(q_{M}e)^2}{t^2}
\end{eqnarray}
Although formally this leads to a divergent total cross-section, the
invariant transfer in our case is simply $2mT$ where $m$ is the mass of the
electrically charged particle and $T$ is the laboratory energy transferred 
to it. The effective cross-section for monopole scattering obtained by
weighing $\frac{d\sigma}{dt}$ with the inelasticity $T/E_{M}= t/2mE_{M}$
is given by
\begin{eqnarray}
\sigma_{eff}= \frac{q^2_{M}e^2}{E_{M}}\int \frac{t}{2m}\frac{1}{t^2} dt
\approx \frac{q^2_{M}e^2}{2mE_{M}}ln(t_{max}/t_{min})
\end{eqnarray}
For $E_{M}$ of order $10^{12}$ GeV, this gives a cross-section for monopole
 scattering of magnitude $10^{-40}$ cm$^2$, which is utterly negligible.

\bigskip
\subsection{Nuclear interaction of monopoles}

The magnetic monopole could have a  hadronic cloud consisting of gluons
and $q\bar{q}$ pairs. This can happen for example if the gauge group
whose breaking leads to the monopoles includes $SU(3)_c$ as a subgroup.
The hadronic cloud will then have a typical dimension $\sim \Lambda^{-1}_{QCD}$
of order one Fermi or so. The hadronic cloud is inherent to the monopole
and is expected to regenerate after every collision. Thus the monopole
 will have a hadronic cross-section for interaction with nuclei similar
to that of a proton. This feature would seem to favor this scenario
for understang the UHE events. In particular,
one event from the Yakutsk experiment appears to be neutron rich
suggesting a hadronic origin for the UHE's.

Unfortunately there is a crucial difference between the inelasticities of 
a monopole initiated and proton initiated UHE reaction. The hadronic outer
layer of the monopole carries only a miniscule fraction $\epsilon\equiv
GeV/ m_{M}$ of the monopole mass. This can be seen for the 't Hooft-Polyakov
monopole, where the bulk of the monopole mass $M_{M}\approx
M_G/g_G$ (where $M_G$ and $g_G$ are the symmetry breaking scales and the
gauge couplings associated with the GUT group whose breaking causes
the appearance of the monopoles) resides in the innermost regions of the
monopole (of dimension $M^{-1}_G$) and an order TeV contribution from the
weak layer of the scale dimension $m^{-1}_W$. The extended hadronic layer
contributes only a fraction $\Lambda_{QCD}/g_s\approx GeV$ to the monopole
mass. In the language of the parton model, the gluons and the $q\bar{q}$
pairs are the "wee" partons in the very 
small x-region of the parton distribution
in the monopole. Since only thses hadronic constituents can participate
in the hadronic monopole nucleus collision, we expect the inelasticity
to be of order $\zeta \leq x\leq GeV/ m_{M}$. This result is clearly
in contrast with the characteristic inelasticities of order $1/2$ in
proton nucleus collision.

Possible binding of the monopole to the nuclei is also not likely to be
important.The reason for this is that the binding potential is given by
\begin{eqnarray}
V=-\frac{{\bf \mu.r}}{r^2} g_{M}
\end{eqnarray}
where ${\bf \mu}$ is the magnetic moment of the nucleus and $g_{M}$
is the magnetic charge of the monopole.  Due to relatively small magnetic
moment of the Oxygen and Nitrogen nuclei, it is not clear whether these
nuclei can sustain the requisite p-wave monopole-nucleus bound state\cite{gold}.
Even if such a bound state formed, it would strip off in the first
nuclear collision and the rate of monopole nuclear capture to reform
the composite is likely to be slower than that of 
the nuclear reactions destroying
it. Hence the mechanism of enhancing nuclear monopole  interaction by
binding them to nuclei and thus enhancing $x_H$ from $GeV/m_{M}$ to 
$x_H\approx A(GeV)/ m_{M}$ 
would be rather inefective. It turns out also that endowing the monopoles
with large cross-sections to catalyze baryonic decays a-la Callan and Rubakov
does not help either\footnote{While this general issue has been
discussed in the past (see e.g.\cite{yyy}), the following argument may
still be worth presenting. Let us consider a blue or red giant star of 
radius $R\approx 3\times (10^{12}-10^{13})$ cm. and column density of
$\approx \frac{10 M_{\odot}}{\pi R^2}\approx 10^8-10^6 gr/cm^2$. Clearly
if the monopoles lose even $10^{-3}$ of their energy in  our atmosphere,
they would stop upon hitting these objects. During its lifetime of
10 million years, the number of monopoles accumulating there is 
$N_M\approx 4\pi^2 \Phi_M R^2 T\approx \Phi_M 10^{42\pm1}$. After the
gravitational collapse to a neutron star, most of these monopoles will
reside in the nuclear core catalyzing one nuclar decay in each hadronic 
time scale of $10^{-23}$ sec.. Altogether in $3\times 10^{16}$ sec.,
each monopole catalyzes $\approx 3\times 10^{39}$ decays destroying a fraction
$10^{-18}$ of the core. To allow the neutron stars to live for
a billion years, we must require that $N_M\leq 10^{18}$ implying that
$\Phi_M\leq 10^{-24\pm1}~cm^{-2}sec^{-1}sr^{-1}$. This would totally
exclude the above scenario.}

Thus in summary, we have shown that massive, stable and neutral colored
particles with masses more than a few hundred GeV are ruled out by
exisiting WIMP searches and masses in the range 100 GeV to few TeV
by the present results on searches for anomalous heavy isotopes. 
They are also not suitable for explanation of the UHE cosmic rays. 
Furthermore, as far as the monopoles are concerned,
while we have not conclusively established that their
 interactions cannot lead to the UHE's, we have raised some
plausible doubts about the effectiveness of the mechanism due to very
small inelasticities in monopole nuclear interaction which can be
inferred from simple intuitive arguments. To clearly confirm or rule
out the mechanism, detailed simulation of the effect of relativistic monopole
air nuclei collision in the various experimental setups is required.
Our net conclusion is that the monopole scenario for the
UHE's is rather unlikely. 

\noindent{\Large \bf{Acknowledgements}}
\bigskip
 The work of R. N. M.
is supported by the National Science Foundation grant no. PHY-9421386
and the work of both R. N. M. and S. N. is supported in part also by
a grant from the Bi-national US-Israel National Science Foundation.



\begin{thebibliography}{99}
                                      
\bibitem{1} A. de Rujula and S. L. Glashow and U. Sarid, Nucl. Phys. 
{\bf B333}, 173 (1990).

S. Dimopoulos, D. Eichler, R. Esmailzadeh and G. Starkman,
 Phys. Rev. {\bf D 41}, 2388 (1990);

A. Gould, B. Draine, R. Romani and S. Nussinov, Phys. Lett. {\bf B 238},
337 (1990).

\bibitem{farrar1} G. Farrar, Phys. Lett. {\bf B265}, 395 (1991); Phys. Rev.
Lett. {\bf 53}, 1029 (184); {\bf 76}, 4111 (1996).

\bibitem{satya} Z. Chacko, B. Dutta, R. N. Mohapatra and S. Nandi, 
hep-ph/9704307.


\bibitem{bere} S. Barr, D. Chang and G. Senjanovi\'c, Phys. Rev. Lett.
{\bf }; Z. Berezhiani and R. N. Mohapatra, Phys. Rev. {\bf D 52}, 6607
(1995); R. Foot and R. Volkas, Phys. Rev. {\bf D 52}, 6595 (1995).

\bibitem{dimo} G. Starkman, A. Gould, R. Esmailzadeh and S. Dimopoulos,
Phys. Rev. {\bf D 41}, 3594 (1990).

\bibitem{hemmick} T. Hemmick et al. Phys. Rev. {\bf D 41}, 2074 (1990).

\bibitem{caldwell} D. Caldwell et al. Phys. Rev. Lett. {\bf 61}, 510 (1988).

\bibitem{avig}  F. Avignone et al. Phys. Rev. {\bf C 34}, 666 (1986).

\bibitem{kuzmin} V. Kuzmin, talk at NANP97, Dubna, July 7-11 (1997);

\bibitem{farrar} D. Chung, G. Farrar and E. Kolb, FERMILAB-PUB-97/187-A;
We thank A. Riotto for pointing out that other massive colored stable
particles may be of interest in this connection.        

\bibitem{nagano} For a recent review, see M. Nagano, Nucl. Phys. B 
(Proc. Suppl.){\bf 52 B}, 71 (1997). 

\bibitem{weil} T. Kephart and T. Weiler, Astropart. Phys. {\bf 4}, 271 (1996).

\bibitem{kolbt} See for instance, E. W. Kolb and M. Turner, {\it Early
Universe}, Addison-Wesley (1989).


\bibitem{nussinov} S. Nussinov, hep-ph/9610236.

\bibitem{kolb} M. Turner, quoted in E. W. Kolb, FERMILAB-Conf.-86/146-A.


\bibitem{elbert} J. Elbert and P. Sommers, Astro-ph/9410069.

\bibitem{XXX} O. Kenneth, private communication.

\bibitem{gold} A. S. Goldhaber, S. Nussinov and L. Stodolsky, Nucl. Phys. 
{\bf B290}, 955 (1988).

\bibitem{yyy} S. Dimopoulos, J. Preskill and F. Wilczek, Phys. Lett. {\bf
B 119}, 320 (1982).

\end{thebibliography}
\end{document}